\documentclass[dvips]{acta}
\usepackage{supertabular,lscape,epsfig}
\usepackage{amssymb}
\usepackage{amsmath}
\usepackage{multirow}
\usepackage[normalem]{ulem}

\begin{document}

\begin{Titlepage}
\Title{Dwarf Nova V1040 Centauri and Variable Stars in its Vicinity}
\Author{A.~Rutkowski$^1$, P.~Pietrukowicz$^2$, A.~Olech$^3$, T.~Ak$^{4}$, K. Z\l{}oczewski$^3$, R.~Poleski$^{2}$, C. Tappert$^{5}$, Z.~Eker$^{1}$}
{{$^1$TUBITAK National Observatory, Akdeniz University Campus, 07058 Antalya, Turkey\\
e-mail:artur@tug.tubitak.gov.tr\\}
{$^2$ Warsaw University Observatory, Al. Ujazdowskie 4, 00-478 Warsaw,
Poland \\ }
{$^3$Nicolaus Copernicus Astronomical Center, Polish Academy of Sciences,\\
Ul. Bartycka 18, 00-716 Warszawa, Poland\\ }
{$^4$Istanbul University, Faculty of Sciences, Department of Astronomy and Space Sciences, 34119 University, Istanbul, Turkey\\ }
{$^5$Departamento de F\'{\i}sica y Astrof\'{\i}sica, Universidad de Valpara\'{\i}so, Av. Gran Breta\~na 1111, Valpara\'{\i}so, Chile}
}
\Received{Month Day, Year}
\end{Titlepage}

\Abstract{We present the results of a photometric campaign of the
dwarf nova V1040 Cen.  The light curve shows two normal outbursts with
recurrence time $\approx$ 40 days and amplitude $\approx$ 2.5 mag.
Quiescence data show oscillations with periods in the range $\approx 0.1$ days
(2.4 h) to $\approx 0.5$ days (12 h) of unknown origin. We measured 
the orbital period of V1040 Cen to be $P_{\rm orb}=0.060458(80)$ days
($1.451\pm0.002$ h). Based on the $M_{\rm v}$--$ P_{\rm orb}$ relation  we
found the distance of V1040 Cen to be $137\pm31$pc. In this paper we also
report the detection of eleven new variable stars in the field of the
monitored dwarf nova.}
{accretion, accretion discs - binaries: cataclysmic variables, stars: dwarfnovae, oscillations, stars: individual: V1040 Cen, RX J1155.4-5641, AAVSO 1150-56, 2MASS J11552726-5641561}

\section{Introduction}
Cataclysmic variables (CVs) are interacting binaries containing a
main-sequence or slightly evolved secondary star losing mass via Roche
lobe overflow onto a white dwarf primary (Warner 1995). In binaries with
a weak magnetic field ($B<10^5$~G) an accretion disc is formed around the
primary. Today it is clear that the accretion disc thermal instability is
the cause of repetitive outbursts observed in some CVs called dwarf
novae (DNe) - see Osaki (1996) for a review.

Dwarf novae present a huge variety of light curve behavior. In this
paper we continue our studies devoted to those stars (\eg Olech et
al. 2009, Rutkowski \etal 2007, 2009). By presenting the analysis of the
light curve of
V1040~Cen we would like to contribute to the ongoing discussion present in the
literature on the origin of superhumps and superoutbursts in SU UMa-type
CVs (\eg Smak 2009, 2010, Rutkowski \etal 2010).

V1040~Cen is a DN with the orbital period of $P_{\rm orb}\approx
0.060416$ days (1.45~h) (Longa-Pena 2009). The distance to this object is
uncertain and is estimated to be 492 pc  by Ak \etal (2008), 
while Pretorius and Knigge (2008b) gives a lower limit distance to this
object as 40 pc.
Superoutburst light curves were previously
studied by Patterson \etal (2003) leading to the discovery of
superhumps with a period of $P_{\rm orb} =  0.06028(10)$ days ($\approx 1.45$~h).
Using high-speed photometry Woudt and Warner (2010) studied Quasi 
Periodic Oscillations (QPOs) and Dwarf Nova Oscillations
(DNOs) in this object. Kato \etal (2003) on the basis of VSNET data
estimated the superoutburst cycle to be $\approx211$~d. Our observations
and its analysis are presented in Sections 2, 3 and 4.

Research devoted to a particular problem also often brings some
by-products in the form of additional discoveries. V1040~Cen is located
close to the Galactic plane in a relatively crowded area. We used this
opportunity to check how the brightness changes for other stars present
in our frames. Careful analysis allowed us to discover eleven new
variable stars which we describe in detail in Section 6.
\section{Observations and Data Reductions}
We observed V1040~Cen using three telescopes at three different
locations: for 4 nights in March 2009 with the 1.0-m Elizabeth
Telescope at the South African Astronomical Observatory (SAAO),
for 31 nights in March-May 2009 with the 1.3-m Warsaw telescope
at Las Campanas Observatory (LCO), and for 14 nights in April 2009
with the SMARTS 0.9-m telescope at Cerro Tololo Inter-American
Observatory (CTIO). Table~1 gives details. 

The richest observation set comes from CTIO where we obtained 2020 frames
in ''white light'' using one $1024\times1024$ quartile of the 2K$\times$2K 
optical imager at a scale of $ 0\zdot\arcs396$/pixel. The exposure times
ranged from 30~s to 90~s. Occasionally we took images in $B$, $V$,
and $I$ filters with exposure times of 300~s, 240~s and 240~s,
respectively.

The data from LCO was taken mostly during the twilight using
a $1024\times2048$ subraster of an eight SITe $2048\times4096$
CCD mosaic camera at a scale of $0\zdot\arcs26$/pixel. Only $V$-band
images with exposure times between 30~s and 90~s were collected.

The SAAO data were collected using back-illuminated CCD camera "STE3" of
a size of $512\times512$~pixels equipped with liquid nitrogen cooling.
Observations were taken in ''white light'' with exposure times of an
order of one minute.

All images were de-biased and flat-fielded using the IRAF\footnote{IRAF
is distributed by the National Optical Astronomy Observatory, which is
operated by the Association of Universities for Research in Astronomy,
Inc., under a cooperative agreement with the National Science
Foundation.} package. The photometry from the CTIO data was extracted
with the help of the {\it Difference Image Analysis Package}
(DIAPL)\footnote{The package is available at
http://users.camk.edu.pl/pych/DIAPL} written by Wo\'zniak (2000) and
recently modified by W. Pych. The package is an implementation of the
method developed by Alard and Lupton (1998). A reference frame was
constructed by combining 13 individual images taken on the night of 2009
April 14/15. Profile photometry for the reference frame was extracted
with DAOPHOT/ALLSTAR (Stetson 1987). These measurements were used to
transform the light curves from differential flux units into
instrumental magnitudes. For the LCO and SAAO data we decided to extract
only aperture photometry with DAOPHOT.

The CTIO images were also used to search for variable stars.
We inspected directly by eye the light curves of 711 stars
in a $6\zdot\arcm2\times6\zdot\arcm2$ field roughly centered
on V1040~Cen. For completeness we performed an independent period
search with the TATRY code (Schwarzenberg-Czerny 1996).

\begin{table}
\centering
\caption{\small Journal of photometric observations of V1040~Cen}
\medskip
{\small
\begin{tabular}{lccr}
\hline
Site         & Observing period     & Filter & Number of frames \\
\hline
SAAO (1.0 m) & 2009 Mar 15 - Mar 18 &   -    &  127 \\
\hline
LCO  (1.3 m) & 2009 Mar 24 - May 3  &  $V$   &  122 \\
\hline
CTIO (0.9 m) & 2009 Apr 2 -15       &   -    & 2020 \\
             &                      &  $B$   &    6 \\
             &                      &  $V$   &    6 \\
             &                      &  $I$   &    6 \\
\hline
\end{tabular}}
\end{table}

\section{Global light curve}
Figure 1 presents the overall light curve from our three-months
observational campaign. Since the filter transmission for the $R$ pass-band is
very close to observations made in "white light" we label the y-axis
of the plot accordingly by $\approx R$ magnitude. 
Three types of symbols indicate data obtained with
 different
telescopes. In this figure one can easily notice three episodes of
significant brightenings. At least two of them (close to HJD $\approx$ 2454906
and HJD $\approx$ 2454946) could be considered as normal dwarf nova
outbursts. Time intervals between consecutive eruptions are $\approx 9$~d and
$\approx31$~d. Assuming that the second brightening (around HJD=2454912.5)
is just fluctuation, the period between two detected  consecutive
outbursts is $\approx40$~d. This value is only slightly different than 
$\approx 35$~d period  mentioned by Woudt and Warner (2010). However,
taking into account the low level amplitude of the first two brightenings and
the small number of points during their peaks we should not draw too
far-reaching conclusions about this period. On the other hand it is also
known that outburst events in CVs are not  strictly periodic so most
likely this is just an effect of this property. 

The $R$ magnitude has been derived as the difference between the
variable and the comparison stars designated by numbers 124, 129 and 141 on
the AAVSO chart. Due to incompatibility of  the photometric systems 
the zero-point errors can likely be as big as 0.2 mag.

\begin{figure}[htb]
\includegraphics[angle=-90, width=\textwidth]{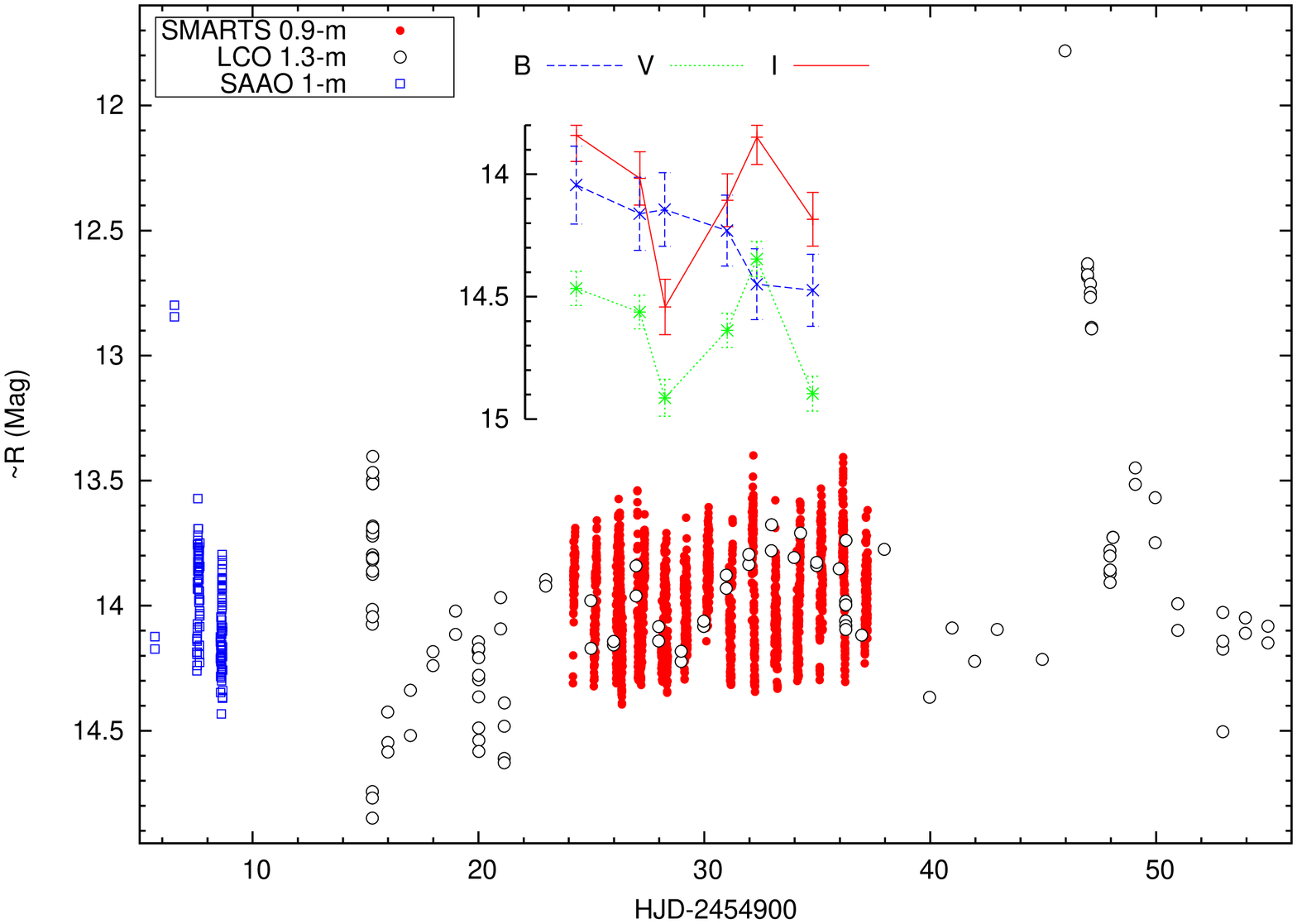}
\FigCap{Complete light curve of V1040 Cen. Dots, open circles and open squares
represent data gathered using SMARTS(CTIO), LCO and SAAO telescopes
 respectively.
The inset presents $BVI$ measurements obtained using the SMART telescope. 
The y-axis of the inset was vertically shifted in order to improve the
readability of the graph. The x-axis scale is the same for the inset and
the rest of the figure.}
\end{figure}

During SMART telescope observations we made occasionally $BVI$
measurements  of  V1040~Cen. The obtained $BVI$ magnitudes are
presented as an inset in Figure 1. These magnitudes were obtained using
comparison stars from AAVSO Variable Star Database and Aladin sky atlas.
Errors come mostly from these transformations. One can clearly see
that $V$ and $I$ magnitudes resemble the LCO light curve.  However the $B$
magnitudes gradually decreases which can be a manifestation of disc
cooling after last outburst. The amplitude of the observed outbursts reaches
2.5 mag.  The duration of the eruptions and other properties
mentioned above strongly suggest, that those were normal dwarf nova
outbursts.
\section{Power spectrum}
\begin{figure}[htb]
\begin{center}
\includegraphics[width=0.8\textwidth]{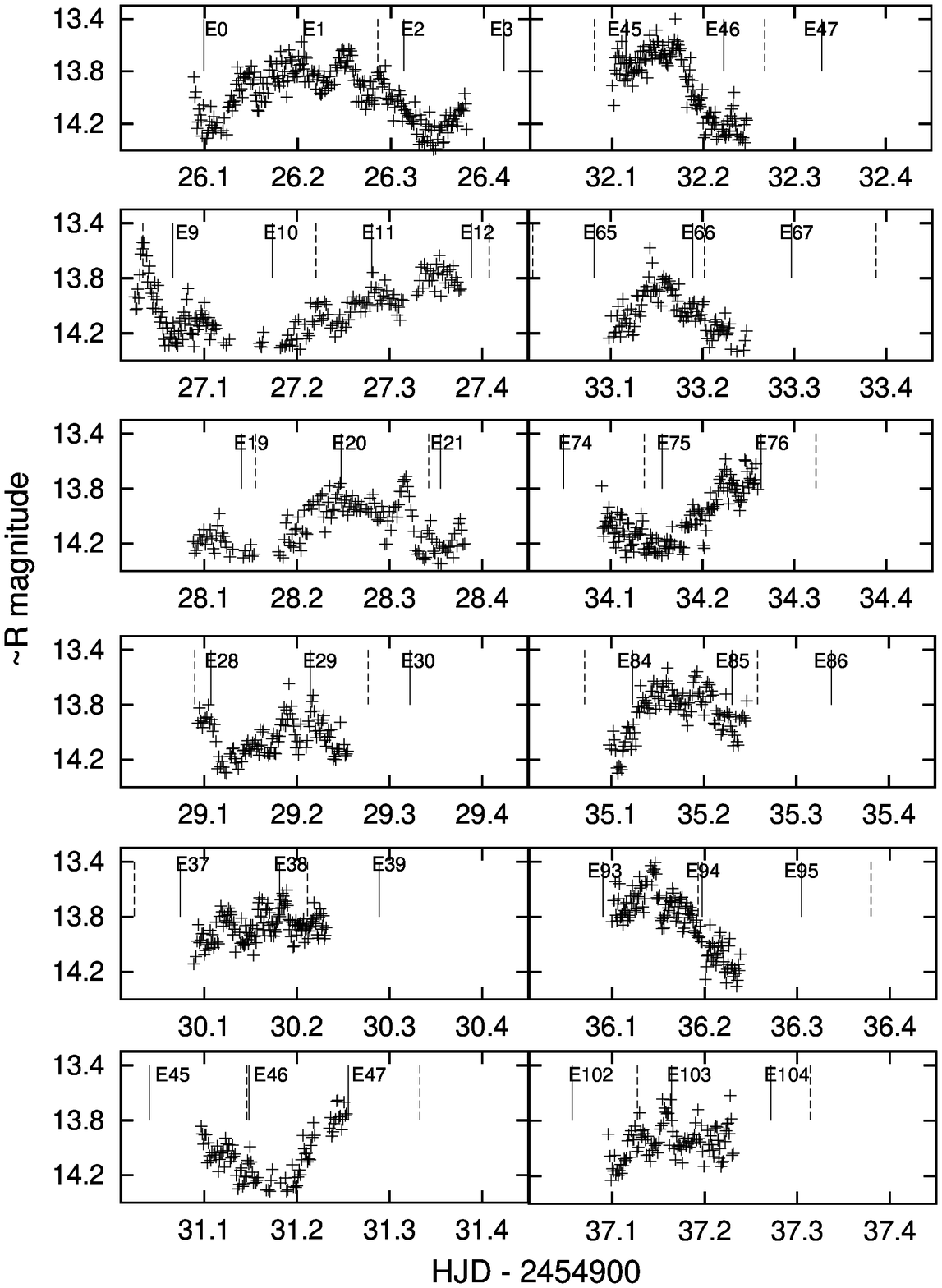}
\end{center}
\FigCap{Example of light curve variations of V1040 Cen obtained with the 0.9-m
SMARTS telescopes. Times of the extrema calculated using the ephemeris
(1) and (2) are also presented. Solid vertical lines indicate timings of 
the light curve minima corresponding to ephemeris (1) while dashed vertical 
lines show timings of the minima corresponding to ephemeris (2).
Only the epoch number of ephemeris (1) is shown for clarity.}
\end{figure}
The data, which span the longest time, were obtained with 1.3-m LCO telescope.
They have, however, quite poor sampling which is not suitable to draw
conclusions about 
variations within timescales of hours. This opportunity is given by SAAO and 
SMARTS data. Figure 2 presents examples of light curve variations obtained 
by the SMARTS telescope. The most prominent variations which can be noticed
have a roughly constant amplitude of $\approx$0.8 mag.

In addition, careful inspection allows us to notice  the smaller amplitude light
variations with an amplitude up to $\approx$0.35 mag superimposed on the
main light curve. 
We use the Fourier transform to reveal all different frequencies that
constitute it. 

The SAAO and SMARTS telescopes data were used for this analysis. The
procedure which we used involves trend removing.  Two approaches
were applied. For each night separately we fitted a first or second 
order polynomial and subtracted it from the original data. This resulted in 
subtracting the most visible $\approx0.8$ mag amplitude variations from the
light curve. Such detrended light curve was analyzed by the ZUZA -
Time Series Analysis code (Schwarzenberg-Czerny 1996). The resulting
periodogram is shown in Figure 3. The highest peak 
at a frequency $f_{\rm orb}=(16.538\pm0.026)\ {\rm c/d}$
corresponds to an orbital period
$P_{\rm orb}=0.060458(80)\ {\rm days}=(1.451\pm0.002)\ {\rm h}$.  
\begin{figure}[htb]
\includegraphics[angle=-90, width=\textwidth]{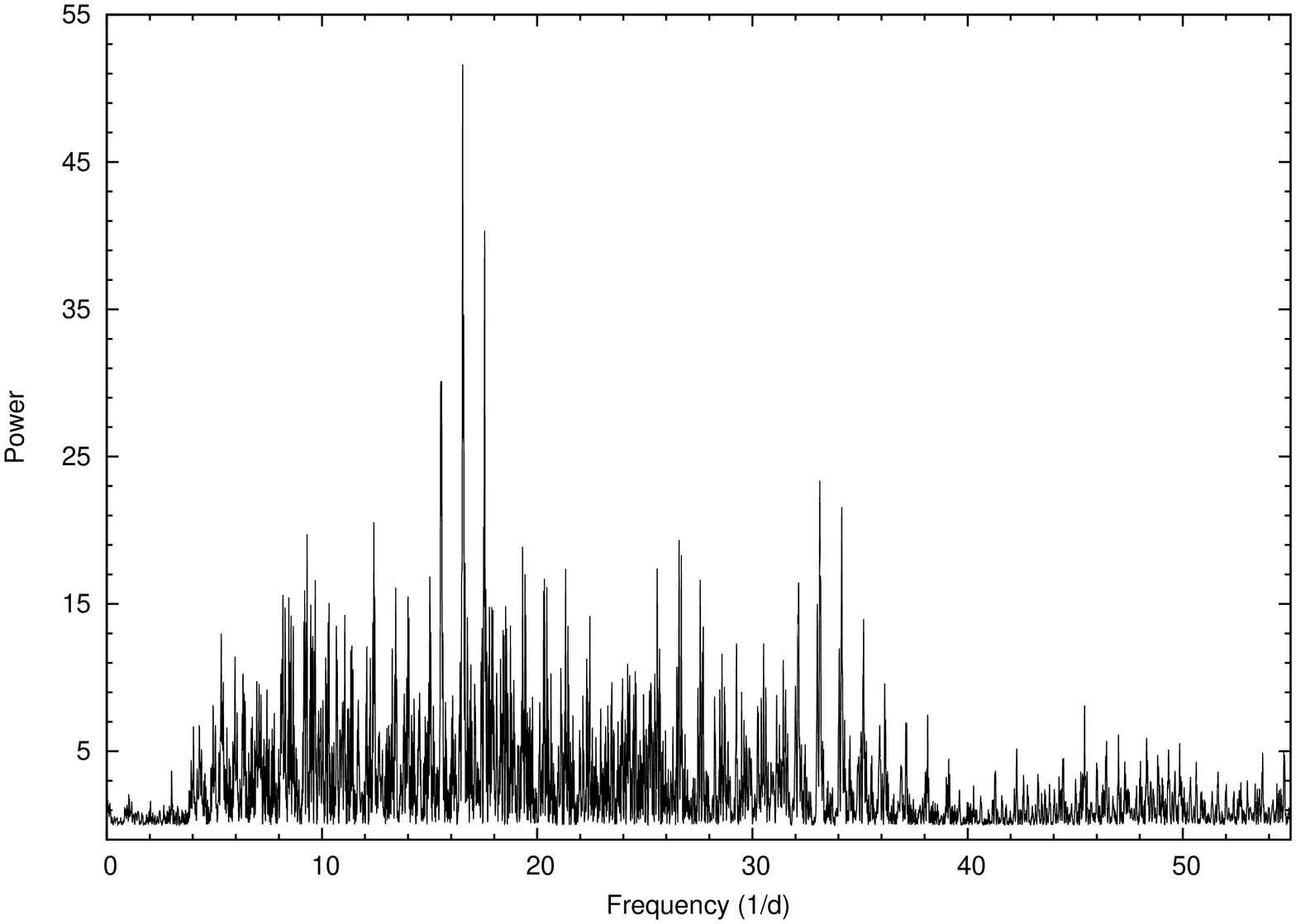}
\FigCap{Periodogram of the data from which a second order polynomial fit has
been removed. The highest pick at $f=16.538$~c/d represents the orbital 
period of the system.}
\end{figure} 
The spectroscopically determined period $(1.45176\pm0.00002)$~h by Longa-Pena (2009)\footnote{http://www.noao.edu/meetings/wildstars2/posters/monday/p-longa-poster.png} falls within this error range.
The second highest peak apart from the daily aliases of $f_{\rm orb}$ 
is $33.1~{\rm c/d} \approx 2f_{\rm orb}$ which we interpret as harmonic
frequency.

As it was mentioned before the most prominent brightness changes have an
amplitude of $\approx0.8$ mag, however this modulation has been removed
during detrending procedure.  Thus we had to repeat the detrending
procedure on the raw data,  but this time we fit to consecutive nights only
first order polynomial. This brought all the curves to the same level
without removing  considered modulations. Next we applied a prewhitening
procedure to remove the orbital frequency $f_{\rm orb}=16.538$~c/d and its
first harmonics from the data. 
The resulting periodogram is presented
in Figure 4. One can see that all  statistically significant
information is present below $f=12$~c/d. Two clear peaks can be noticed --
namely $f_{1}=5.35$~c/d and $f_{2}=9.31$~c/d. After investigation of the
periodogram one can notice that the remaining variations are not strictly
periodic -- there are several minor peaks spread around the frequencies 
$f_1$ and $f_2$  within the range $\pm \approx2$~c/d. 

\begin{figure}[htb]
\includegraphics[angle=-90, width=\textwidth]{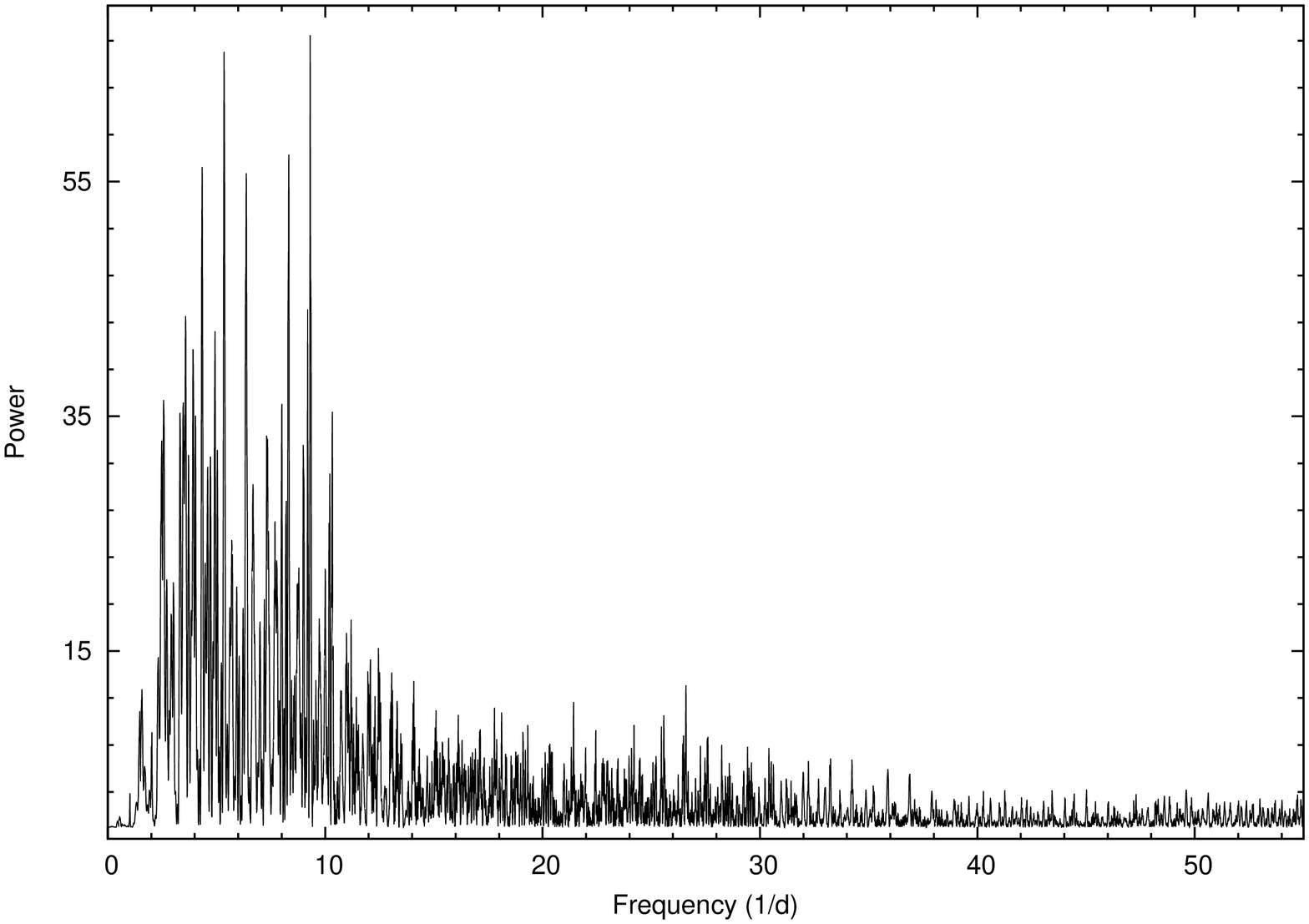}
\FigCap{Power spectrum obtained for the data from which a first order polynomial
fit has been subtracted. The orbital frequency is also subtracted. The two
highest peaks likely
correspond to modulations present in the light curve. (Compare with Fig. 2 and Fig. 5)}
\end{figure}

In order to check if the detected oscillation period varies randomly or
rather shows systematic lengthening or shortening we performed an $O-C$
analysis. To construct it we
 used the periods corresponding to frequencies $f_1$ and $f_2$. 
Since in most cases minima are  much sharper and better
visible we used them for this analysis. The results, however, are
inconclusive and difficult to interpret.  There is no noticeable
tendency in obtained $O-C$ diagram.  For better
visualization of this behavior we put vertical markers corresponding to the
timings of minima in the Fig. 2. The markers locations were
calculated based on the following ephemeris:
\begin{equation}
E1_{min} = 26.14(1) + 0^{\rm d}.1863(2)\cdot E
\end{equation}

\noindent for the dashed lines, and

\begin{equation}
E2_{min} = 26.1138(52) + 0^{\rm d}.10750(7)\cdot E
\end{equation}

\noindent for the solid lines.
The resulting light curves are shown in Fig 5.
\begin{figure}[htb]
\includegraphics[width=\textwidth]{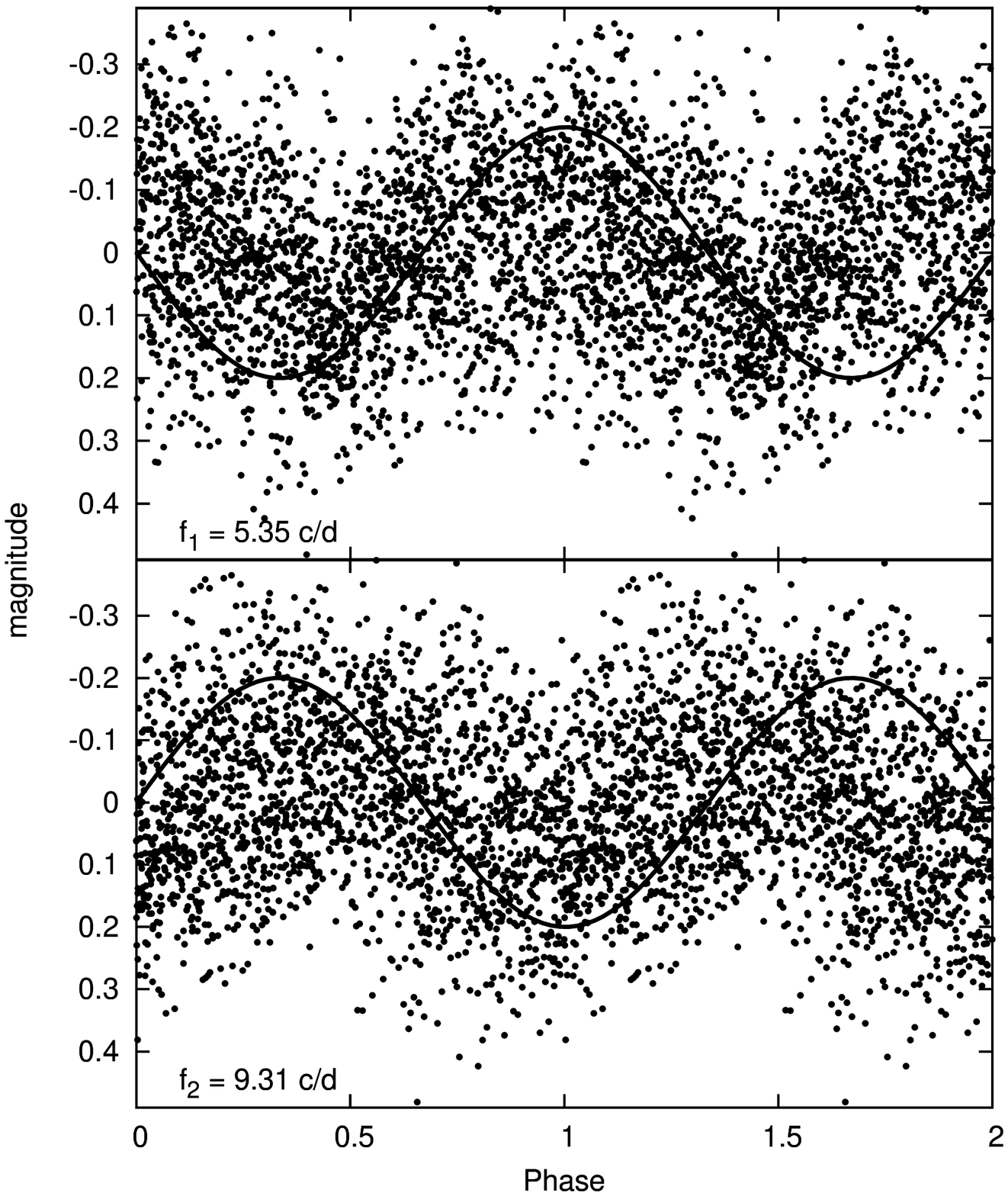}
\FigCap{The combined detrended data of V1040 Cen folded on the frequencies
$f_1=5.35$~c/d and $f_2=9.31$~c/d on top and bottom panel respectively.}
\end{figure}

At least in the case of two other objects (\ie SDSS J162520.29 - Olech et
al. 2011, and 1RXS J053234.9 - Rutkowski \etal 2010) the similar low frequency
signals below $\approx10$~c/d were also observed. So far their nature  is
unknown. The presence of this kind 
of low frequency oscillations (for dwarf novae inside or below the period gap)
is puzzling.  

One can speculate that explanation of this low frequency signals
can involve oscillations of the hot spot and/or reflect the
long-term variation of the mass transfer rate.

\section{Distance to V1040 Cen}
So far, there are two estimates of the distance of V1040 Cen. 
Pretorius and Knigge (2008b) estimated a lower limit of 40 pc for the distance
to the system. Their distance determination method was based on infrared
magnitudes that are calculated from theoretical models of the secondary
stars and orbital periods by Knigge (2006) (see also, Pretorius and
Knigge, 2008a). They assumed that the secondary stars fill their Roche lobe.
However, Ak \etal's (2007, 2008) period-luminosity-colour (PLC) relation gives a
value of 492 pc for the lower limit of distance to V1040 Cen. 
%

The difference between two estimates of distance to V1040 Cen comes from the
absolute magnitudes calculated by Ak \etal (2008) and Knigge (2006).
Ak \etal (2008) found the system's absolute brightness in $JHK_{\rm s}$
 photometric bands as twice the theoretical $JHK_{\rm s}$ absolute 
brightness found by Knigge (2006). 

A different approach is to use the outburst maximum brightness in the $V$ band.
This procedure is based on the assumption that
the accretion disc maximum brightness during outburst of a dwarf nova is a
good standard candle. Based on the relation given by Warner (1995, eq 3.4):

\begin{equation}
M_{\rm v}({\rm max}) = 5.74 -0.259 P_{\rm orb} {\rm [h]}
\end{equation}

Using $P_{\rm orb} = 1.45$~h, for V1040 Cen we obtain $M_{\rm
v}=(5.36\pm0.2)$~mag. The brightness of V1040 Cen during the observed
outburst maximum was measured to be $(11.48\pm0.2)$~mag.  Based on
Schlegel, Finkbeiner \& Davis (1998) exctinction maps at NASA/IPAC 
Infrared Science
Archive\footnote{http://irsa.ipac.caltech.edu/applications/DUST/} for
the position of V1040 Cen we obtain an extinction $A_{\rm v}=1.38$~mag.
Using the
above numbers we were able to estimate the distance to V1040 Cen to be
${\rm d}=(88\pm31)$ pc, where the error was calculated based on exact
differential and corresponds to 1 sigma. 
This means that the 40 pc of Pretorius \& Knigge (2008b)
- which additionally is a lower limit - is well within 2 sigmas of our
value.
Moreover, one can notice that this 40 pc value was
based on the semi-empirical donor sequence.
So, one might expect these lower limits presented in their work are typically
underestimated true distances by about a factor of 2.
Therefore distance estimation of Pretorius \& Knigge (2008b) of V1040 Cen 
would actually be around 80 pc, which is in excellent agreement 
with our $88 \pm 31$ pc estimate.

On the other hand, value of 492 pc obtained by Ak \etal (2008)
significantly disagree with our result. 
But one should keep in mind that
the Schlegel's $A_{\rm v}$ absorption value, is the total absoption from
us to the edge of the Galaxy - therefore the absorption for V1040 Cen might
be lower.
If we take $E(B-V) = 0.139$ mag from Ak \etal  (2008) then 
we obtain $A_{\rm v}=0.431$ mag. Consequently a new distance estimate
is about $137\pm 31$ pc. This new corrected distance value, however, still
remain in disagreement with 492 pc given by Ak \etal (2008) but is well within
2 sigmas with dubled lower limit estimations ($\approx 80$ pc) of
Pretorius \& Knigge (2008b). However, we believe that $A_{\rm v}$ inferred
from Ak \etal (2008) is correct and, therefore we take ${\rm d}=137 \pm 31$ pc
as the most likely value.

In addition if main source of the errors comes from inaccurate determination  of
the outburst maximum, then, to achieve agreement with Ak \etal (2008)
${m_v}$ must be different by at least $\approx 2$~mag
which seems unlikely. However one should take into account that the
accretion disk brightness depends strongly on the viewing angle. 
Paczy\'nski and Schwarzenberg-Czerny (1980) analysed this effect 
and presented a formula which shows that equation (3) works fine only
for intermediate ($\approx 57^{\rm o}$) system inclinations. Therefore, 
the disk will be $\approx 1$~mag brighter for face-on
systems or $\approx 1.7$~mag
faither for close to edge-on systems.
\section{Variable stars in the field of V1040 Cen}
V1040 Cen is located 5.3~deg from the Galactic plane in a relatively
crowded area. The 14-day-long observations at CTIO allowed us to search
for variable objects in the vicinity of the monitored dwarf nova. Within the
$6\zdot\arcm2\times6\zdot\arcm2$ field we found eleven new variables with
brightness between 14.80 and 19.65~mag in the $R$ band. In Table~2 we give
information on equatorial coordinates, brightness, possible period and
type of each variable. In Figs.~6 and 7 we present light curves and finding 
charts, respectively.

\begin{table}
\centering
\caption{\small Basic data on detected variables in the field of V1040 Cen.}
\medskip
{\small
\begin{tabular}{lccccll}
\hline
Var & RA(2000.0)  & Dec(2000.0) & $R_{\rm max}$ & $A_R$ & $P$ & Type \\
    &             &             &     [mag]     & [mag] & [d] & \\
\hline
V1  & 11:55:06.34 & -56:43:30.1 & 15.67 & 0.02 & 7.5(5)     & puls? \\
V2  & 11:55:06.79 & -56:42:12.8 & 14.98 & 0.04 &            & per? \\
V3  & 11:55:08.43 & -56:40:35.5 & 16.84 & 0.64 & 0.41215(5) & EW \\
V4  & 11:55:10.57 & -56:40.49.6 & 17.37 & 0.11 &            & per? \\
V5  & 11:55:18.50 & -56:40:07.2 & 17.50 & 0.03 &            & per? \\
V6  & 11:55:21.74 & -56:42:29.0 & 19.65 & 0.80 & 0.5859(1)  & EA \\
V7  & 11:55:24.97 & -56:39:18.7 & 14.80 & 0.03 & 9.1(1)     & puls? \\
V8  & 11:55:26.77 & -56:42:26.7 & 16.11 & 0.06 & 8.9(1)     & puls? \\
V9  & 11:55:42.86 & -56:41:35.1 & 17.13 & 0.50 & 0.7137(1)  & EB \\
V10 & 11:55:45.77 & -56:41:23.8 & 18.95 & 0.25 & 2.510(1)   & EA \\
V11 & 11:55:46.93 & -56:41:09.4 & 15.62 & 0.04 & 1.27(1)    & ecl? \\

\hline
\end{tabular}}
\end{table}

\begin{figure}[htb]
\includegraphics[angle=0, width=\textwidth]{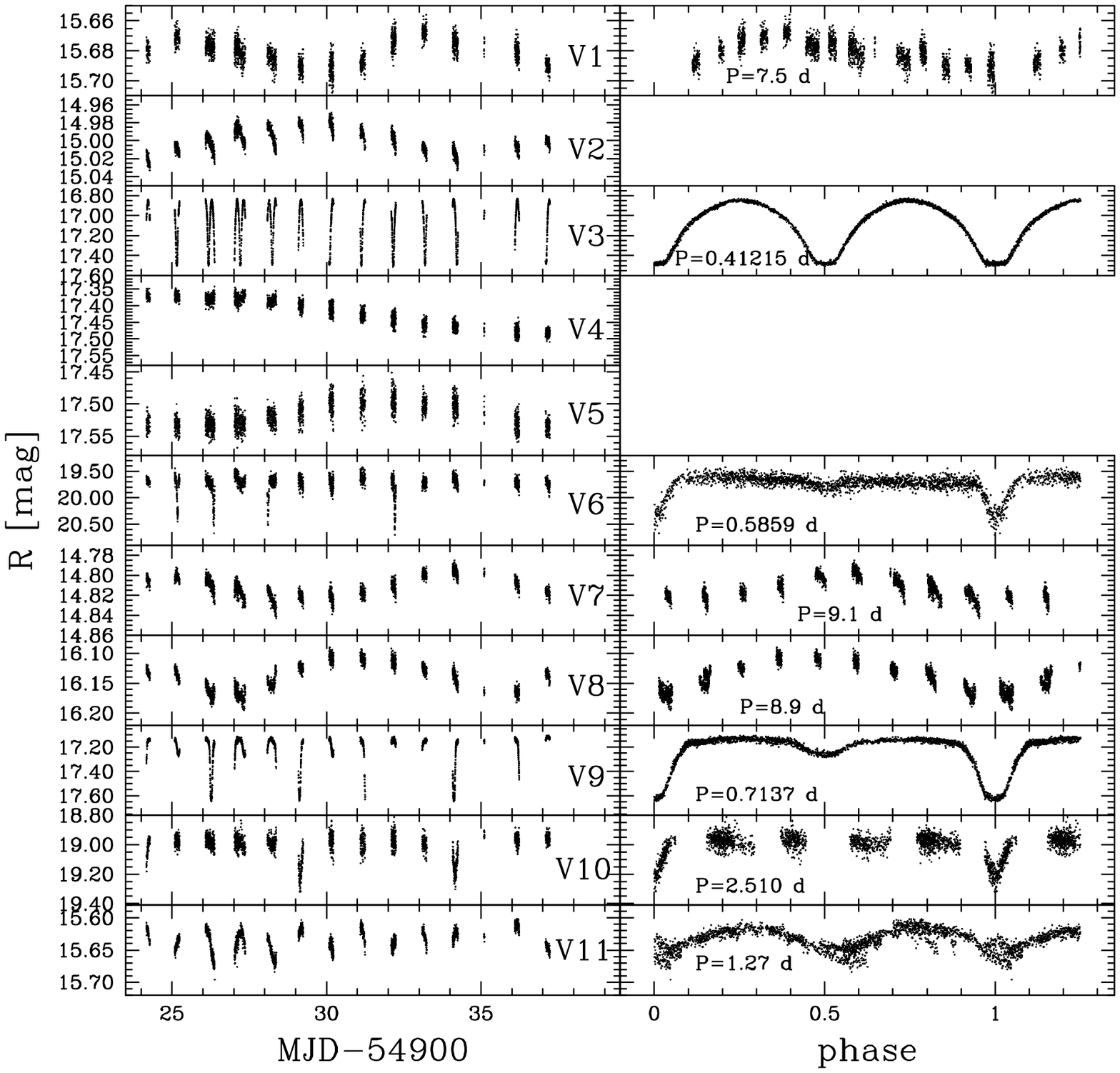}
\FigCap{Time-domain (left column) and phased (right column) light curves
of eleven newly discovered variable stars in the field of V1040~Cen.}
\end{figure}

\begin{figure}[htb]
\includegraphics[angle=0, width=\textwidth]{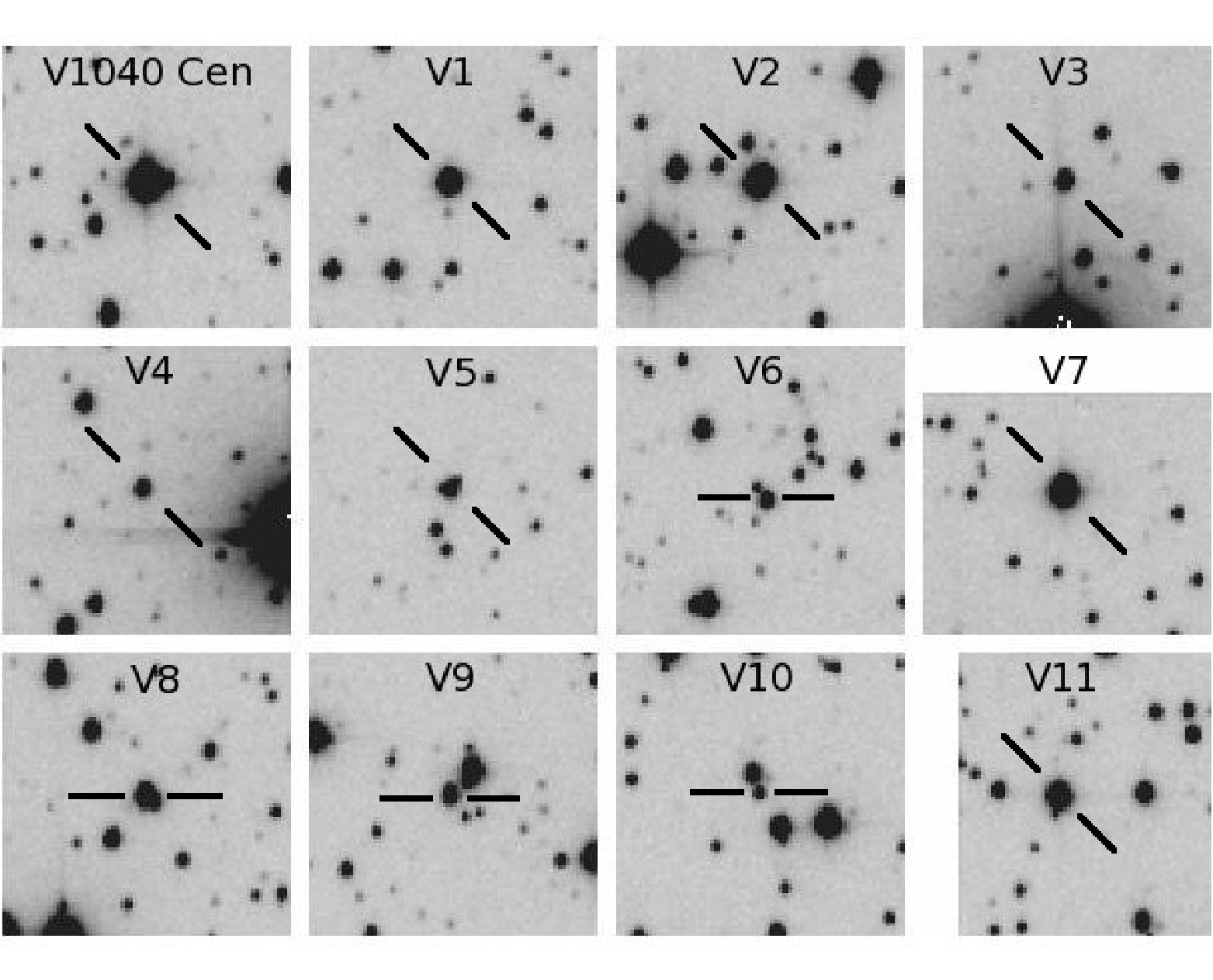}
\FigCap{Finding charts for V1040~Cen and eleven newly discovered
variables. Each chart is 32 arcsec on a side. North is up
and East is to the left.}
\end{figure}

Four of the objects, V3, V6, V9, and V10, are {\it bona fide} eclipsing
systems. We note that the orbital period of V10 can be twice longer than the
given value of 2.510~d. The variable V3 is a W~UMa-type binary star showing flat
minima of the same depth of 0.64~mag. 

The nature of the other seven objects is not clear. Only for V1, V7, V8, and V11
we were able to derive periods. Long-period variables, namely V1, V7, and V8,
are probably pulsating stars, while V11 seems to be an eclipsing binary.
More observations are needed to definitively answer the question what type
of objects they are.

\section{Conclusions}
\begin{itemize}
\item[a)] We have presented the results of the observational campaign of the V1040 Cen. Light
curve shows at least two normal outbursts. The $\approx 40$ days
gap between them fits well with recurrence time period estimated
 by Woudt \& Warner (2010).
$\approx 40$ days and amplitude reaching 2.5 mag.
\item[b)] The quiescence data show oscillations with the period range from 
$\approx 2.4$h to $\approx 4.8$h  which can not be easily linked
to any known phenomenon.
\item[c)] We have found the orbital period of V1040 Cyg to be 
$P_{\rm orb}=0.060458(80)$~days ($1.451\pm0.002$~h), which is also 
in excellent agreement with the orbital period determined from spectroscopy. 
\item[d)] We estimated the distance to V1040 Cen to be ($137\pm31$) pc. 
\item[e)] We detected eleven new variables in the field of V1040~Cen.
Four of them are eclipsing binaries.

\end{itemize}

\Acknow{This work was partly supported by the Polish MNiSW grant
No. N203~301~335 to AO. AR is supported by 2221-Visiting Scientist
Fellowship Program of TUBITAK. PP and RP is supported from funding to the
OGLE project from the European Research Council under the European
Community$'$s Seventh Framework Programme (FP7/2007-2013)/ERC grant
agreement no. 246678. PP is also supported by the grant
 No. IP2010 031570 financed by the Polish Ministry of Sciences and Higher
 Education under Iuventus Plus programme.}

\end{document}